\documentclass[reprint,amsmath,amssymb,braket,aps,prb]{revtex4-2}
\usepackage{graphicx,color,braket}
\usepackage{dcolumn}
\usepackage{bm}
\usepackage[colorlinks=true,linkcolor=blue]{hyperref}
\usepackage{colortbl}
\begin{document}

\title{Response times of two-dimensional photodetectors limited by intrinsic resistance and capacitance}

\author{Ilya Safonov$^{1,2}$ and Dmitry Svintsov$^{1}$}
\affiliation{$^{1}$ Laboratory of 2d Materials for Optoelectronics, Center for Photonics and 2D Materials, Moscow Institute of Physics and Technology, Dolgoprudny, 141700, Russia}
\email{svintcov.da@mipt.ru}
\affiliation{$^{2}$ Programmable Functional Materials Lab, Center for Neurophysics and Neuromorphic Technologies, Moscow 127495, Russia}

\begin{abstract}
Abstract. Most contemporary architectures of photodetectors based on two-dimensional materials include global gates for carrier density control and local p-n junctions in the channel. We study the dependence of photocurrent in such detectors on the light modulation frequency, fully taking into account the effects of distributed resistance and gate-channel capacitance. The decay of photocurrent with modulation frequency governs the response time. We find that the maximum modulation frequency is largely determined by the position of light-sensitive junction with respect to the middle of the channel. Largest modulation frequency is achieved for junctions in immediate vicinity of either source or drain contacts, while fast roll-off of the modulation characteristic is observed for junction in the middle of the channel. 
\end{abstract}

\maketitle

\section{Introduction}

Photodetectors based on two-dimensional (2d) materials feature ultrafast response and possibility of tuning via electrical gating~\cite{Rogalski2019a}. Response times can be divided into intrinsic times governing the formation of local photocurrent, and extrinsic times governing the formation of voltage and current in an electronic circuit. The intrinsic times in most 2d photodetectors are very short. In graphene, where photovoltage is generated via thermoelectric effect, the intrinsic time is that of carrier thermalization \cite{Tielrooij2015}. The latter is estimated as hundreds of femtoseconds. In 2d electron systems based on quantum wells, photocurrent generation via plasmon drag is often dominant \cite{Kurita2014}. The intrinsic time of such effect is that of momentum relaxation, order of 1…10 picoseconds in high-quality heterostructures~\cite{Muravev2016}.

Short intrinsic photoresponse implies that it can be easily masked by circuit effects, i.e. by resistive and capacitive delays. The truly external delays are minimized by matching the photodetectors with coplanar lines. However, many photodetectors have their own channel resistance and internal capacitances: the capacitance between source and drain contacts, and capacitance of the channel with respect to the gate, as shown in Fig. 1A. The distributed character of resistance and capacitance in 2d detectors makes the question about RC-limited response time complex. Despite the existence of several circuit models of gated photodetectors~\cite{Rudin2015,coquillat2016high}, it is not yet understood which resistance and capacitance govern the response time and maximum modulation frequency.

Currently, there exist two main approaches for radiation detection with 2d materials. The first one lies in the electrical induction of n- and p-type regions within the channel via application of opposite voltages to the split gates~\cite{Castilla2020,Titova2023a,Soundarapandian2024}. The gate-induced p-n junction is generally located in the middle of the channel, and the photocurrent $j_{ph}(x)$ is generated locally at the junction, Fig. 1B. The second approach lies in deposition of metal contacts with dissimilar work function~\cite{Koepfli2023,Wei2021}, Fig. 1C. Carrier diffusion between metals and 2d channel leads to the formation of p-i-n doping profile. The photocurrent $j_{ph}(x)$ in such architecture is generated in the immediate vicinity of metal leads. In some rare situations, like bulk photovoltaic~\cite{Otteneder2020} or photon drag effects~\cite{Karch2010}, the photocurrent is generated in 2d channel uniformly, Fig. 1D.

We study the response time of gated 2d photodetectors vs. their structural parameters: channel resistance, channel length, gate-channel capacitance and position of the light-sensitive junction. We theoretically consider the photocurrent response Iph to the radiation with modulation frequency $f_{mod}$ and show that $I_{ph}(f_{mod})$ drops with frequency. We show that the maximum frequency depends strongly on the position of light-sensitive junction with respect to source and drain contacts, and find that structures with photocurrent generation in immediate vicinity of metal leads are superior to those with junction in the mid-channel.

\begin{figure}[ht]
	\centering
	\includegraphics[width=1.0\linewidth]{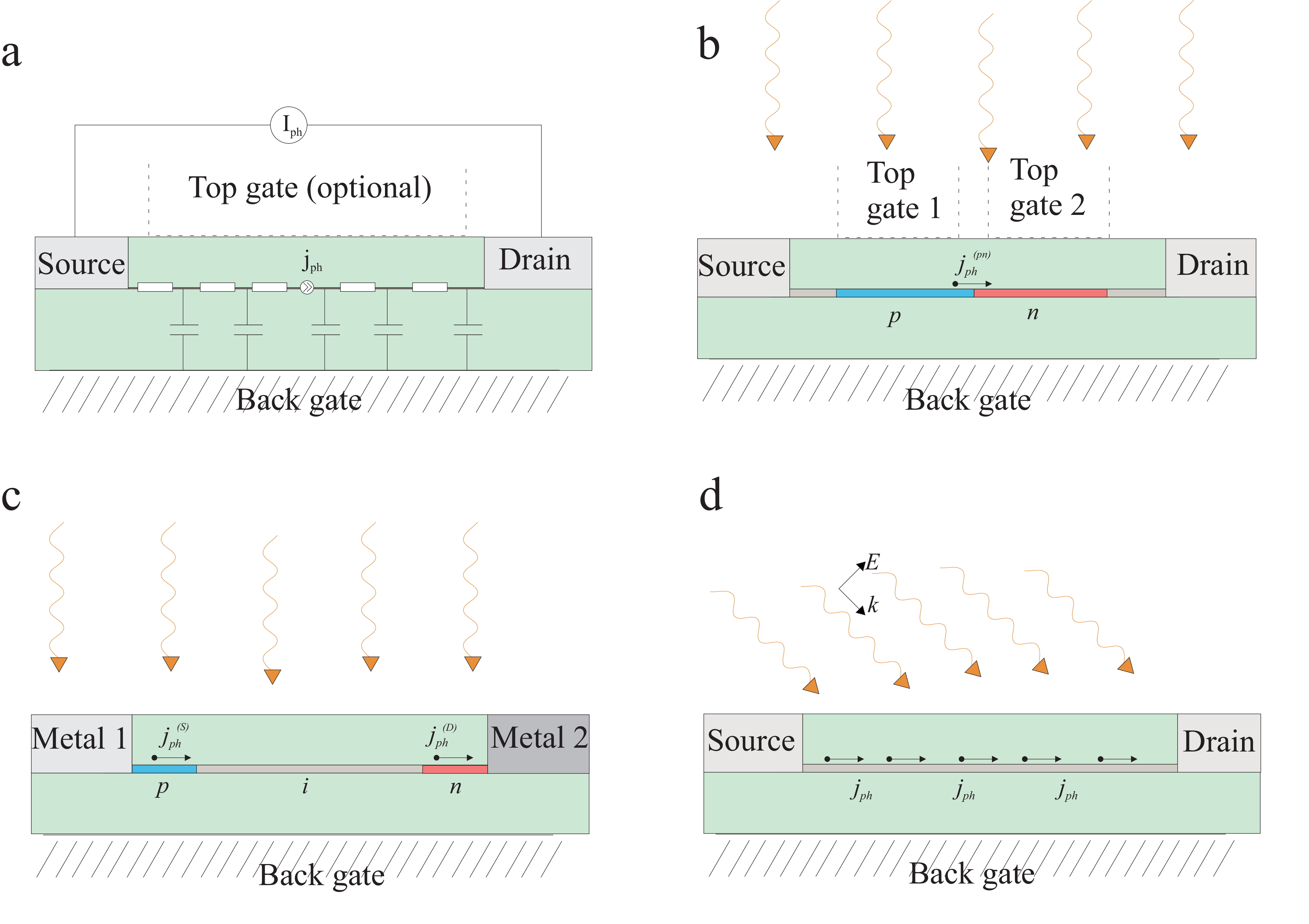}
	\caption{(A) Schematic of the gated photodetector with 2d channel and its equivalent distributed RC circuit with localized photocurrent source (B-D) Possible physical realizations of 2d photodetectors fitting the scheme of panel (A): detector with p-n junction induced by split gate (B), detector with p-i-n junction due to dissimilar metal contacts (C), detector based on photon drag effect (D)}
	\label{Fig-strcuture}
\end{figure}

\section{Methods}
 
We consider the photoresponse of 2d detector with channel of length $L$, source and drain contacts located at $x=0$ and $x=L$, respectively. The detector is globally gated, and the gate-channel separation is $d\ll L$. The emerging photocurrent $I_{ph}$ is measured between source and drain contacts. We model $I_{ph}$ at the circuit level without specifying the microscopic origin of the photoresponse. More precisely, we present the overall current density $j\left( x \right)$ as a sum of photocurrent ${{j}_{ph}}\left( x \right)$ and drift current $-\sigma dV/dx$, where $\sigma$ is the conductivity of 2d material and $V$is the local voltage in the channel:
\begin{equation}
j\left( x \right)={{j}_{ph}}\left( x \right)-\sigma \frac{dV}{dx}.
\end{equation}

The role of drift current is especially clear for localized distribution of photocurrent density, i.e. for characteristic width of the function ${{j}_{ph}}\left( x \right)$ well below the source-drain separation $L$. The local electric field $dV/dx$ in this case builds up in the regions where is ${{j}_{ph}}\left( x \right)$ zero, such that the total current becomes spatially uniform. For time-independent illumination, requiring the spatial uniformity of the current, we find 
\begin{equation}
	{{I}_{ph,dc}}=\frac{1}{L}\int\limits_{0}^{L}{{{j}_{ph}}\left( x \right)dx},
\end{equation}
i.e. the global photocurrent is the spatial average of local photocurrents.
The photoresponse to the modulated radiation is far less trivial. We consider the harmonic variations of local photocurrent ${{j}_{ph}}\left( x,t \right)={{j}_{ph}}\left( x \right){{e}^{-i\omega t}}$, where $\omega =2\pi {{f}_{\bmod }}$ is the modulation frequency. The buildup of local electric potential is now not instantaneous and governed by the continuity equation:
\begin{equation}
-i\omega \rho \left( x \right)+\frac{d}{dx}\left( {{j}_{ph}}\left( x \right)-\sigma \frac{dV}{dx} \right)=0,	
\end{equation}
where $\rho \left( x \right)$ is the local charge density in the channel. We link the charge density $\rho \left( x \right)$ and electric potential $V\left( x \right)$ via the Poisson’s equation. For 2d channels with proximate gates, the Poisson’s equation can is solved in the local capacitance approximation:
\begin{equation}
\rho \left( x \right)={{C}_{A}}\left[ V\left( x \right)-{{V}_{G}} \right],	
\end{equation}
where ${{C}_{A}}=\varepsilon {{\varepsilon }_{0}}/d$is the gate-to-channel capacitance per unit area, and ${{V}_{G}}$ is the gate voltage. Considering the smallness of photo induced voltage and potential as compared to their dc values, we write $\rho \left( x \right)={{C}_{A}}V\left( x \right).$ The final equation for distribution of photoinduced potential becomes
\begin{equation}
V\left( x \right)+\frac{1}{q_{p}^{2}}\frac{{{d}^{2}}V\left( x \right)}{d{{x}^{2}}}=\frac{1}{i\omega {{C}_{A}}}\frac{d{{j}_{ph}}\left( x \right)}{dx};	
\end{equation}

In the regime of photocurrent measurement its boundary conditions are $V\left( 0 \right)=V\left( L \right)=0$. Above, ${{q}_{p}}=\sqrt{i\omega {{C}_{A}}/\sigma }$ is the complex parameter governing the spatial propagation of electric potential. In the low-frequency regime, the conductivity $\sigma \equiv {{\sigma }_{dc}}$ is real, and the parameter ${{q}_{p}}$ is complex, ${{q}_{p}}={{e}^{i\pi /4}}\sqrt{\omega {{C}_{A}}/{{\sigma }_{dc}}}$. In such situations the spreading of potential is diffusive. Similar equation has been adopted in earlier papers~\cite{Dyakonov1996,Ryzhii2006} for the analysis of plasmon-assisted resonant photoresponse of detectors.

Determination of photocurrent $I_{ph}$ in external circuit for time-dependent response is based on Shockley-Ramo theorem. It states that $I_{ph}$ is the weighted average of local currents
\begin{equation}
{{I}_{ph}}=\int\limits_{0}^{L}{g\left( x \right)j\left( x \right)dx}	
\end{equation}
where the weight function $g\left( x \right)$is the electric field in the channel created by unity voltage between source and drain. In the case of proximate gates considered here, the source-drain electric field is strongly screened by the gate, and persists only in the immediate vicinity of lateral contacts, $g\left( x \right)\approx \left[ \delta \left( x \right)+\delta \left( x-L \right) \right]/2$~\cite{Svintsov2018b}. The current measured Iph in such case is the average of local currents at the source and drain:
\begin{equation}
{{I}_{ph}}=-\sigma \left( {{\left. \frac{dV}{dx} \right|}_{x=0}}+{{\left. \frac{dV}{dx} \right|}_{x=L}} \right).	
\end{equation}
Equation (5) is readily solved for arbitrary spatial dependence of ${{j}_{ph}}\left( x \right)$, and current in external circuit is symbolically presented as 
\begin{equation}
{{I}_{ph}}=\frac{1}{2}\int\limits_{0}^{L}{\frac{\sin \left[ {{q}_{p}}x \right]-\sin \left[ {{q}_{p}}\left( L-x \right) \right]}{\sin \left[ {{q}_{p}}L \right]}\frac{d{{j}_{ph}}}{dx}dx}.	
\end{equation}
The frequency dependence of photocurrent is fully encoded in parameter ${{q}_{p}}\left( \omega  \right)={{e}^{i\pi /4}}\sqrt{\omega {{C}_{A}}/{{\sigma }_{dc}}}$. The wave vector ${{q}_{p}}\left( \omega  \right)$ can be presented as ${{q}_{p}}\left( \omega  \right)={{e}^{i\pi /4}}\sqrt{\omega RC}/L$, where R is the total channel resistance and C is the total gate-channel capacitance. Thus, Eq. (8) generalizes the concept of RC-delay to the detectors with distributed resistance and capacitance.

\section{Results and Discussion}
To highlight the dependence of response time and maximum modulation frequency on localization of light-sensitive junction, we adopt the box-shaped model of photocurrent ${{j}_{ph}}\left( x \right)=j_{ph}^{\left( 0 \right)}\theta \left[ x-\left( {{x}_{ph}}-w/2 \right) \right]\theta \left[ \left( {{x}_{ph}}+w/2 \right)-x \right]$. Physically, the light-sensitive junction has length $w$, centered at ${{x}_{ph}}$, and generates constant photocurrent $j_{ph}^{\left( 0 \right)}$. The model describes all three setups of Fig. 1 B-D via appropriate choice of ${{x}_{ph}}$ and $w$. Plugging the model photocurrent distribution into expression (8), we find the current in external circuit:
\begin{multline}
{{I}_{ph}}\left( \omega  \right)=\\
j_{ph}^{\left( 0 \right)}\sin \left( \frac{{{q}_{p}}w}{2} \right)\left[ \cos \left( {{q}_{p}}{{x}_{ph}} \right)\cot \left( \frac{{{q}_{p}}L}{2} \right)+\sin \left( {{q}_{p}}{{x}_{ph}} \right) \right].	
\end{multline}
Equation (9) is our central result. Figure 2 shows the computed dependences of $I_{ph}$ on modulation frequency at various junction positions ${{x}_{ph}}$ and small junction width $w=L/20$. The model parameters are adequate to contemporary graphene detectors, $L=10$ $\mu$m, $d=100$ nm, $\sigma^{ -1}=100$ Ohm, $\varepsilon$=4. We observe that the frequency roll-off of the photocurrent is fastest when the junction is positioned in the middle of the channel, and becomes slower as the junction gets closer to the source. The dependence ${{I}_{ph}}\left( \omega  \right)$ is symmetric with respect to the mid-channel, i.e. the frequency roll-off is identical for ${{x}_{ph}}={{x}_{0}}$ and ${{x}_{ph}}=L-{{x}_{0}}$. The computed dependence supports our intuition about the propagation of photovoltage signals in distributed photodetectors. If the photocurrent is generated away from the source and drain, it takes a long time for the signal propagation to the contacts, therefore the maximum modulation frequency is low. On the contrary, for signal generated in immediate vicnity of either contact, the RC-limited response time is very short. Only a small fraction of channel resistance and gate-channel capacitance contributes to the apparent $R$ and $C$. For very small separations between junction and the contact (case ${{x}_{ph}}=L/5$ in Fig. 2), the photocurrent can even initially grow with frequency. 

\begin{figure}[ht]
	\centering
	\includegraphics[width=1.0\linewidth]{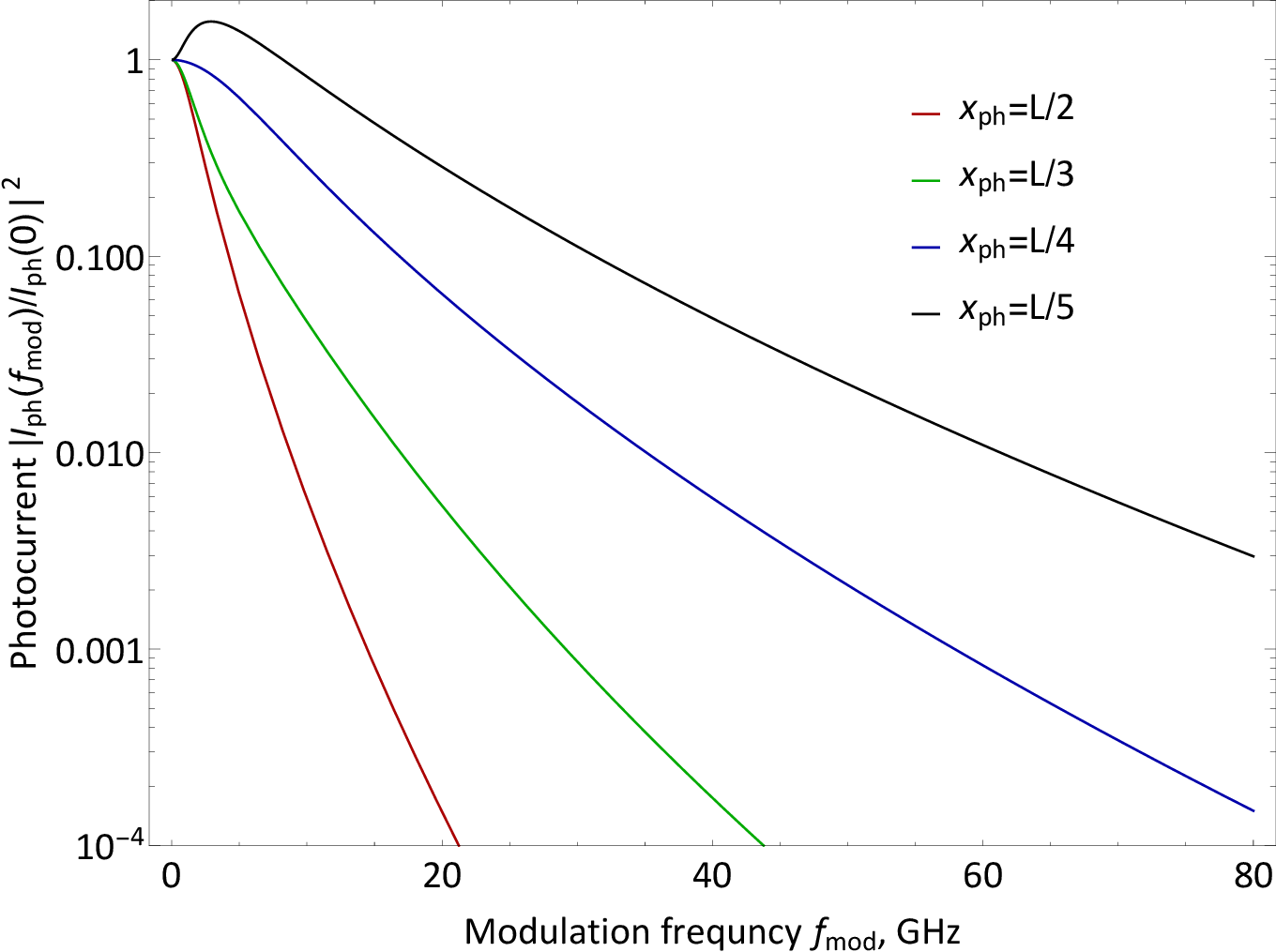}
	\caption{Dependence of photocurrent on modulation frequency ${{f}_{\bmod }}=\omega /2\pi $ computed with Eq. (9) for different positions of the light-sensitive junction ${{x}_{ph}}$. Detector parameters are channel length L=10 µm, gate-channel separation $d=100$ nm, gate dielectric constant $\varepsilon =4$, channel resistivity $\sigma -1=100$ Ohm}
	\label{Fig-result}
\end{figure}

At very high frequencies the photocurrent eventually decays according to the asymptotic law
\begin{equation}
	{{I}_{ph}}\left( \omega  \right)=j_{ph}^{\left( 0 \right)}\frac{{{q}_{p}}w}{2}\exp \left( -\frac{\left| {{q}_{p}} \right|}{\sqrt{2}}\min \left\{ {{x}_{ph}},L-{{x}_{ph}} \right\} \right).
\end{equation}
The maximum modulation frequency $f_{\bmod }^{\max }$, i.e. the frequency at which the photodetector response drops by e times, is given by
\begin{equation}
f_{ mod }^{\max }=\frac{1}{\pi RC}\frac{L}{\min \left\{ {{x}_{ph}},L-{{x}_{ph}} \right\}}.	
\end{equation}
An intuitive interpretation of (11) is the following: only a fraction of total resistance and capacitance located between the light-sensitive junction metal contact is relevant to the RC-delay of the distributed photodetector.

Finally, we discuss the limitations of our model. First of all, it is based on local capacitance approximation which is valid for proximate gates $\left| {{q}_{p}} \right|d<<1$. This assumption is best fulfilled at low frequencies, particularly, for the parameters of Fig. 2 and ${{f}_{\bmod }}=10$ GHz we estimate ${{\left| {{q}_{p}} \right|}^{-1}}\approx 700$ nm, which is well above the distance to the gate. Attempting to go beyond the local capacitance approximation, one can replace Eq. (4) with the full Poisson’s equation taking into account realistic geometry of the source, drain and gate contacts. For certain simple arrangements of contacts, it can be solved with conformal mapping techniques~\cite{Ryzhii1995}.

Another approximation of our model is the harmonic approximation to the modulated local photocurrent ${{j}_{ph}}\left( x,t \right)={{j}_{ph}}\left( x \right){{e}^{-i\omega t}}$. Such approximation can be valid if only the modulation frequency is well below the radiation frequency, ${f}_{mod}\ll {{f}_{rad}}$. The assumption can be restrictive for sub-terahertz detectors considered for future wireless communications, where ${{f}_{\text{rad}}}=100...300$ GHz. The applicability of our model is limited to tens of gigahertz in that case. For optical fiber detectors ($\lambda =1.55$ $\mu$m, ${{f}_{\text{rad}}}\approx 200$ THz), the applicability of the model is very broad and definitely covers the contemporary achievable modulation frequencies.

\section{Conclusion}
We have generalized the concept of RC delay to the photodetectors based on 2d materials with global gates and local light-sensitive p-n junctions. We have shown that ‘proper’ resistance and capacitance governing the maximum modulation frequency and the response time are those of the part of the channel between light-sensitive p-n junction and most proximate contact (either source or drain). We have derived a general expression for the frequency roll-off of the photoresponse valid at arbitrary light modulation frequencies and arbitrary distribution of the photocurrent within the channel.

\section{Acknowledgments}
The work was supported by the grant 21-79-20225-P of the Russian Science Foundation. The author acknowledge helpful discussions with Viacheslav Muravev on the technology of high-frequency measurements of photodetectors.

\bibliography{Refs}

\begin{thebibliography}{17}%
\makeatletter
\providecommand \@ifxundefined [1]{%
 \@ifx{#1\undefined}
}%
\providecommand \@ifnum [1]{%
 \ifnum #1\expandafter \@firstoftwo
 \else \expandafter \@secondoftwo
 \fi
}%
\providecommand \@ifx [1]{%
 \ifx #1\expandafter \@firstoftwo
 \else \expandafter \@secondoftwo
 \fi
}%
\providecommand \natexlab [1]{#1}%
\providecommand \enquote  [1]{``#1''}%
\providecommand \bibnamefont  [1]{#1}%
\providecommand \bibfnamefont [1]{#1}%
\providecommand \citenamefont [1]{#1}%
\providecommand \href@noop [0]{\@secondoftwo}%
\providecommand \href [0]{\begingroup \@sanitize@url \@href}%
\providecommand \@href[1]{\@@startlink{#1}\@@href}%
\providecommand \@@href[1]{\endgroup#1\@@endlink}%
\providecommand \@sanitize@url [0]{\catcode `\\12\catcode `\$12\catcode
  `\&12\catcode `\#12\catcode `\^12\catcode `\_12\catcode `\%12\relax}%
\providecommand \@@startlink[1]{}%
\providecommand \@@endlink[0]{}%
\providecommand \url  [0]{\begingroup\@sanitize@url \@url }%
\providecommand \@url [1]{\endgroup\@href {#1}{\urlprefix }}%
\providecommand \urlprefix  [0]{URL }%
\providecommand \Eprint [0]{\href }%
\providecommand \doibase [0]{https://doi.org/}%
\providecommand \selectlanguage [0]{\@gobble}%
\providecommand \bibinfo  [0]{\@secondoftwo}%
\providecommand \bibfield  [0]{\@secondoftwo}%
\providecommand \translation [1]{[#1]}%
\providecommand \BibitemOpen [0]{}%
\providecommand \bibitemStop [0]{}%
\providecommand \bibitemNoStop [0]{.\EOS\space}%
\providecommand \EOS [0]{\spacefactor3000\relax}%
\providecommand \BibitemShut  [1]{\csname bibitem#1\endcsname}%
\let\auto@bib@innerbib\@empty
\bibitem [{\citenamefont {Rogalski}\ \emph {et~al.}(2019)\citenamefont
  {Rogalski}, \citenamefont {Kopytko},\ and\ \citenamefont
  {Martyniuk}}]{Rogalski2019a}%
  \BibitemOpen
  \bibfield  {author} {\bibinfo {author} {\bibfnamefont {A.}~\bibnamefont
  {Rogalski}}, \bibinfo {author} {\bibfnamefont {M.}~\bibnamefont {Kopytko}},\
  and\ \bibinfo {author} {\bibfnamefont {P.}~\bibnamefont {Martyniuk}},\
  }\bibfield  {title} {\bibinfo {title} {{Two-dimensional infrared and
  terahertz detectors: Outlook and status}},\ }\href
  {https://doi.org/10.1063/1.5088578} {\bibfield  {journal} {\bibinfo
  {journal} {Applied Physics Reviews}\ }\textbf {\bibinfo {volume} {6}},\
  \bibinfo {pages} {021316} (\bibinfo {year} {2019})}\BibitemShut {NoStop}%
\bibitem [{\citenamefont {Tielrooij}\ \emph {et~al.}(2015)\citenamefont
  {Tielrooij}, \citenamefont {Piatkowski}, \citenamefont {Massicotte},
  \citenamefont {Woessner}, \citenamefont {Ma}, \citenamefont {Lee},
  \citenamefont {Myhro}, \citenamefont {Lau}, \citenamefont {Jarillo-Herrero},
  \citenamefont {van Hulst},\ and\ \citenamefont {Koppens}}]{Tielrooij2015}%
  \BibitemOpen
  \bibfield  {author} {\bibinfo {author} {\bibfnamefont {K.~J.}\ \bibnamefont
  {Tielrooij}}, \bibinfo {author} {\bibfnamefont {L.}~\bibnamefont
  {Piatkowski}}, \bibinfo {author} {\bibfnamefont {M.}~\bibnamefont
  {Massicotte}}, \bibinfo {author} {\bibfnamefont {A.}~\bibnamefont
  {Woessner}}, \bibinfo {author} {\bibfnamefont {Q.}~\bibnamefont {Ma}},
  \bibinfo {author} {\bibfnamefont {Y.}~\bibnamefont {Lee}}, \bibinfo {author}
  {\bibfnamefont {K.~S.}\ \bibnamefont {Myhro}}, \bibinfo {author}
  {\bibfnamefont {C.~N.}\ \bibnamefont {Lau}}, \bibinfo {author} {\bibfnamefont
  {P.}~\bibnamefont {Jarillo-Herrero}}, \bibinfo {author} {\bibfnamefont
  {N.~F.}\ \bibnamefont {van Hulst}},\ and\ \bibinfo {author} {\bibfnamefont
  {F.~H.~L.}\ \bibnamefont {Koppens}},\ }\bibfield  {title} {\bibinfo {title}
  {{Generation of photovoltage in graphene on a femtosecond timescale through
  efficient carrier heating}},\ }\href {https://doi.org/10.1038/nnano.2015.54}
  {\bibfield  {journal} {\bibinfo  {journal} {Nature Nanotechnology}\ }\textbf
  {\bibinfo {volume} {10}},\ \bibinfo {pages} {437} (\bibinfo {year} {2015})},\
  \Eprint {https://arxiv.org/abs/1504.06487} {arXiv:1504.06487} \BibitemShut
  {NoStop}%
\bibitem [{\citenamefont {Kurita}\ \emph {et~al.}(2014)\citenamefont {Kurita},
  \citenamefont {Ducournau}, \citenamefont {Coquillat}, \citenamefont {Satou},
  \citenamefont {Kobayashi}, \citenamefont {{Boubanga Tombet}}, \citenamefont
  {Meziani}, \citenamefont {Popov}, \citenamefont {Knap}, \citenamefont
  {Suemitsu},\ and\ \citenamefont {Otsuji}}]{Kurita2014}%
  \BibitemOpen
  \bibfield  {author} {\bibinfo {author} {\bibfnamefont {Y.}~\bibnamefont
  {Kurita}}, \bibinfo {author} {\bibfnamefont {G.}~\bibnamefont {Ducournau}},
  \bibinfo {author} {\bibfnamefont {D.}~\bibnamefont {Coquillat}}, \bibinfo
  {author} {\bibfnamefont {A.}~\bibnamefont {Satou}}, \bibinfo {author}
  {\bibfnamefont {K.}~\bibnamefont {Kobayashi}}, \bibinfo {author}
  {\bibfnamefont {S.}~\bibnamefont {{Boubanga Tombet}}}, \bibinfo {author}
  {\bibfnamefont {Y.~M.}\ \bibnamefont {Meziani}}, \bibinfo {author}
  {\bibfnamefont {V.~V.}\ \bibnamefont {Popov}}, \bibinfo {author}
  {\bibfnamefont {W.}~\bibnamefont {Knap}}, \bibinfo {author} {\bibfnamefont
  {T.}~\bibnamefont {Suemitsu}},\ and\ \bibinfo {author} {\bibfnamefont
  {T.}~\bibnamefont {Otsuji}},\ }\bibfield  {title} {\bibinfo {title}
  {{Ultrahigh sensitive sub-terahertz detection by InP-based asymmetric
  dual-grating-gate high-electron-mobility transistors and their broadband
  characteristics}},\ }\href {https://doi.org/10.1063/1.4885499} {\bibfield
  {journal} {\bibinfo  {journal} {Applied Physics Letters}\ }\textbf {\bibinfo
  {volume} {104}},\ \bibinfo {pages} {251114} (\bibinfo {year}
  {2014})}\BibitemShut {NoStop}%
\bibitem [{\citenamefont {Muravev}\ \emph {et~al.}(2016)\citenamefont
  {Muravev}, \citenamefont {Solov'ev}, \citenamefont {Fortunatov},
  \citenamefont {Tsydynzhapov},\ and\ \citenamefont {Kukushkin}}]{Muravev2016}%
  \BibitemOpen
  \bibfield  {author} {\bibinfo {author} {\bibfnamefont {V.~M.}\ \bibnamefont
  {Muravev}}, \bibinfo {author} {\bibfnamefont {V.~V.}\ \bibnamefont
  {Solov'ev}}, \bibinfo {author} {\bibfnamefont {A.~A.}\ \bibnamefont
  {Fortunatov}}, \bibinfo {author} {\bibfnamefont {G.~E.}\ \bibnamefont
  {Tsydynzhapov}},\ and\ \bibinfo {author} {\bibfnamefont {I.~V.}\ \bibnamefont
  {Kukushkin}},\ }\bibfield  {title} {\bibinfo {title} {{On the response time
  of plasmonic terahertz detectors}},\ }\href
  {https://doi.org/10.1134/S0021364016120080} {\bibfield  {journal} {\bibinfo
  {journal} {JETP Letters}\ }\textbf {\bibinfo {volume} {103}},\ \bibinfo
  {pages} {792} (\bibinfo {year} {2016})}\BibitemShut {NoStop}%
\bibitem [{\citenamefont {Rudin}\ \emph {et~al.}(2015)\citenamefont {Rudin},
  \citenamefont {Rupper},\ and\ \citenamefont {Shur}}]{Rudin2015}%
  \BibitemOpen
  \bibfield  {author} {\bibinfo {author} {\bibfnamefont {S.}~\bibnamefont
  {Rudin}}, \bibinfo {author} {\bibfnamefont {G.}~\bibnamefont {Rupper}},\ and\
  \bibinfo {author} {\bibfnamefont {M.}~\bibnamefont {Shur}},\ }\bibfield
  {title} {\bibinfo {title} {{Ultimate response time of high electron mobility
  transistors}},\ }\href {https://doi.org/10.1063/1.4919706} {\bibfield
  {journal} {\bibinfo  {journal} {Journal of Applied Physics}\ }\textbf
  {\bibinfo {volume} {117}},\ \bibinfo {pages} {174502} (\bibinfo {year}
  {2015})}\BibitemShut {NoStop}%
\bibitem [{\citenamefont {Coquillat}\ \emph {et~al.}(2016)\citenamefont
  {Coquillat}, \citenamefont {Nodjiadjim}, \citenamefont {Blin}, \citenamefont
  {Konczykowska}, \citenamefont {Dyakonova}, \citenamefont {Consejo},
  \citenamefont {Nouvel}, \citenamefont {P{\`e}narier}, \citenamefont {Torres},
  \citenamefont {But} \emph {et~al.}}]{coquillat2016high}%
  \BibitemOpen
  \bibfield  {author} {\bibinfo {author} {\bibfnamefont {D.}~\bibnamefont
  {Coquillat}}, \bibinfo {author} {\bibfnamefont {V.}~\bibnamefont
  {Nodjiadjim}}, \bibinfo {author} {\bibfnamefont {S.}~\bibnamefont {Blin}},
  \bibinfo {author} {\bibfnamefont {A.}~\bibnamefont {Konczykowska}}, \bibinfo
  {author} {\bibfnamefont {N.}~\bibnamefont {Dyakonova}}, \bibinfo {author}
  {\bibfnamefont {C.}~\bibnamefont {Consejo}}, \bibinfo {author} {\bibfnamefont
  {P.}~\bibnamefont {Nouvel}}, \bibinfo {author} {\bibfnamefont
  {A.}~\bibnamefont {P{\`e}narier}}, \bibinfo {author} {\bibfnamefont
  {J.}~\bibnamefont {Torres}}, \bibinfo {author} {\bibfnamefont
  {D.}~\bibnamefont {But}}, \emph {et~al.},\ }\bibfield  {title} {\bibinfo
  {title} {High-speed room temperature terahertz detectors based on inp double
  heterojunction bipolar transistors},\ }\href@noop {} {\bibfield  {journal}
  {\bibinfo  {journal} {International Journal of High Speed Electronics and
  Systems}\ }\textbf {\bibinfo {volume} {25}},\ \bibinfo {pages} {1640011}
  (\bibinfo {year} {2016})}\BibitemShut {NoStop}%
\bibitem [{\citenamefont {Castilla}\ \emph {et~al.}(2020)\citenamefont
  {Castilla}, \citenamefont {Vangelidis}, \citenamefont {Pusapati},
  \citenamefont {Goldstein}, \citenamefont {Autore}, \citenamefont
  {Slipchenko}, \citenamefont {Rajendran}, \citenamefont {Kim}, \citenamefont
  {Watanabe}, \citenamefont {Taniguchi}, \citenamefont {Mart{\'{i}}n-Moreno},
  \citenamefont {Englund}, \citenamefont {Tielrooij}, \citenamefont
  {Hillenbrand}, \citenamefont {Lidorikis},\ and\ \citenamefont
  {Koppens}}]{Castilla2020}%
  \BibitemOpen
  \bibfield  {author} {\bibinfo {author} {\bibfnamefont {S.}~\bibnamefont
  {Castilla}}, \bibinfo {author} {\bibfnamefont {I.}~\bibnamefont
  {Vangelidis}}, \bibinfo {author} {\bibfnamefont {V.-V.}\ \bibnamefont
  {Pusapati}}, \bibinfo {author} {\bibfnamefont {J.}~\bibnamefont {Goldstein}},
  \bibinfo {author} {\bibfnamefont {M.}~\bibnamefont {Autore}}, \bibinfo
  {author} {\bibfnamefont {T.}~\bibnamefont {Slipchenko}}, \bibinfo {author}
  {\bibfnamefont {K.}~\bibnamefont {Rajendran}}, \bibinfo {author}
  {\bibfnamefont {S.}~\bibnamefont {Kim}}, \bibinfo {author} {\bibfnamefont
  {K.}~\bibnamefont {Watanabe}}, \bibinfo {author} {\bibfnamefont
  {T.}~\bibnamefont {Taniguchi}}, \bibinfo {author} {\bibfnamefont
  {L.}~\bibnamefont {Mart{\'{i}}n-Moreno}}, \bibinfo {author} {\bibfnamefont
  {D.}~\bibnamefont {Englund}}, \bibinfo {author} {\bibfnamefont {K.-J.}\
  \bibnamefont {Tielrooij}}, \bibinfo {author} {\bibfnamefont {R.}~\bibnamefont
  {Hillenbrand}}, \bibinfo {author} {\bibfnamefont {E.}~\bibnamefont
  {Lidorikis}},\ and\ \bibinfo {author} {\bibfnamefont {F.~H.~L.}\ \bibnamefont
  {Koppens}},\ }\bibfield  {title} {\bibinfo {title} {{Plasmonic antenna
  coupling to hyperbolic phonon-polaritons for sensitive and fast mid-infrared
  photodetection with graphene}},\ }\href
  {https://doi.org/10.1038/s41467-020-18544-z} {\bibfield  {journal} {\bibinfo
  {journal} {Nature Communications}\ }\textbf {\bibinfo {volume} {11}},\
  \bibinfo {pages} {4872} (\bibinfo {year} {2020})},\ \Eprint
  {https://arxiv.org/abs/2006.00358} {arXiv:2006.00358} \BibitemShut {NoStop}%
\bibitem [{\citenamefont {Titova}\ \emph {et~al.}(2023)\citenamefont {Titova},
  \citenamefont {Mylnikov}, \citenamefont {Kashchenko}, \citenamefont
  {Safonov}, \citenamefont {Zhukov}, \citenamefont {Dzhikirba}, \citenamefont
  {Novoselov}, \citenamefont {Bandurin}, \citenamefont {Alymov},\ and\
  \citenamefont {Svintsov}}]{Titova2023a}%
  \BibitemOpen
  \bibfield  {author} {\bibinfo {author} {\bibfnamefont {E.}~\bibnamefont
  {Titova}}, \bibinfo {author} {\bibfnamefont {D.}~\bibnamefont {Mylnikov}},
  \bibinfo {author} {\bibfnamefont {M.}~\bibnamefont {Kashchenko}}, \bibinfo
  {author} {\bibfnamefont {I.}~\bibnamefont {Safonov}}, \bibinfo {author}
  {\bibfnamefont {S.}~\bibnamefont {Zhukov}}, \bibinfo {author} {\bibfnamefont
  {K.}~\bibnamefont {Dzhikirba}}, \bibinfo {author} {\bibfnamefont {K.~S.}\
  \bibnamefont {Novoselov}}, \bibinfo {author} {\bibfnamefont {D.~A.}\
  \bibnamefont {Bandurin}}, \bibinfo {author} {\bibfnamefont {G.}~\bibnamefont
  {Alymov}},\ and\ \bibinfo {author} {\bibfnamefont {D.}~\bibnamefont
  {Svintsov}},\ }\bibfield  {title} {\bibinfo {title} {{Ultralow-noise
  Terahertz Detection by p-n Junctions in Gapped Bilayer Graphene}},\ }\href
  {https://doi.org/10.1021/acsnano.2c12285} {\bibfield  {journal} {\bibinfo
  {journal} {ACS Nano}\ }\textbf {\bibinfo {volume} {17}},\ \bibinfo {pages}
  {8223} (\bibinfo {year} {2023})},\ \Eprint {https://arxiv.org/abs/2212.05352}
  {arXiv:2212.05352} \BibitemShut {NoStop}%
\bibitem [{\citenamefont {Soundarapandian}\ \emph {et~al.}(2024)\citenamefont
  {Soundarapandian}, \citenamefont {Castilla}, \citenamefont {Koepfli},
  \citenamefont {Marconi}, \citenamefont {Kulmer}, \citenamefont {Vangelidis},
  \citenamefont {de~la Bastida}, \citenamefont {Rongione}, \citenamefont
  {Tongay}, \citenamefont {Watanabe}, \citenamefont {Taniguchi}, \citenamefont
  {Lidorikis}, \citenamefont {Tielrooij}, \citenamefont {Leuthold},\ and\
  \citenamefont {Koppens}}]{Soundarapandian2024}%
  \BibitemOpen
  \bibfield  {author} {\bibinfo {author} {\bibfnamefont {K.~P.}\ \bibnamefont
  {Soundarapandian}}, \bibinfo {author} {\bibfnamefont {S.}~\bibnamefont
  {Castilla}}, \bibinfo {author} {\bibfnamefont {S.~M.}\ \bibnamefont
  {Koepfli}}, \bibinfo {author} {\bibfnamefont {S.}~\bibnamefont {Marconi}},
  \bibinfo {author} {\bibfnamefont {L.}~\bibnamefont {Kulmer}}, \bibinfo
  {author} {\bibfnamefont {I.}~\bibnamefont {Vangelidis}}, \bibinfo {author}
  {\bibfnamefont {R.}~\bibnamefont {de~la Bastida}}, \bibinfo {author}
  {\bibfnamefont {E.}~\bibnamefont {Rongione}}, \bibinfo {author}
  {\bibfnamefont {S.}~\bibnamefont {Tongay}}, \bibinfo {author} {\bibfnamefont
  {K.}~\bibnamefont {Watanabe}}, \bibinfo {author} {\bibfnamefont
  {T.}~\bibnamefont {Taniguchi}}, \bibinfo {author} {\bibfnamefont
  {E.}~\bibnamefont {Lidorikis}}, \bibinfo {author} {\bibfnamefont {K.-J.}\
  \bibnamefont {Tielrooij}}, \bibinfo {author} {\bibfnamefont {J.}~\bibnamefont
  {Leuthold}},\ and\ \bibinfo {author} {\bibfnamefont {F.~H.~L.}\ \bibnamefont
  {Koppens}},\ }\bibfield  {title} {\bibinfo {title} {{High-Speed
  Graphene-based Sub-Terahertz Receivers enabling Wireless Communications for
  6G and Beyond}},\ }\href {http://arxiv.org/abs/2411.02269} {\ ,\ \bibinfo
  {pages} {1} (\bibinfo {year} {2024})},\ \Eprint
  {https://arxiv.org/abs/2411.02269} {arXiv:2411.02269} \BibitemShut {NoStop}%
\bibitem [{\citenamefont {Koepfli}\ \emph {et~al.}(2023)\citenamefont
  {Koepfli}, \citenamefont {Baumann}, \citenamefont {Koyaz}, \citenamefont
  {Gadola}, \citenamefont {G{\"{u}}ng{\"{o}}r}, \citenamefont {Keller},
  \citenamefont {Horst}, \citenamefont {Nashashibi}, \citenamefont
  {Schwanninger}, \citenamefont {Doderer}, \citenamefont {Passerini},
  \citenamefont {Fedoryshyn},\ and\ \citenamefont {Leuthold}}]{Koepfli2023}%
  \BibitemOpen
  \bibfield  {author} {\bibinfo {author} {\bibfnamefont {S.~M.}\ \bibnamefont
  {Koepfli}}, \bibinfo {author} {\bibfnamefont {M.}~\bibnamefont {Baumann}},
  \bibinfo {author} {\bibfnamefont {Y.}~\bibnamefont {Koyaz}}, \bibinfo
  {author} {\bibfnamefont {R.}~\bibnamefont {Gadola}}, \bibinfo {author}
  {\bibfnamefont {A.}~\bibnamefont {G{\"{u}}ng{\"{o}}r}}, \bibinfo {author}
  {\bibfnamefont {K.}~\bibnamefont {Keller}}, \bibinfo {author} {\bibfnamefont
  {Y.}~\bibnamefont {Horst}}, \bibinfo {author} {\bibfnamefont
  {S.}~\bibnamefont {Nashashibi}}, \bibinfo {author} {\bibfnamefont
  {R.}~\bibnamefont {Schwanninger}}, \bibinfo {author} {\bibfnamefont
  {M.}~\bibnamefont {Doderer}}, \bibinfo {author} {\bibfnamefont
  {E.}~\bibnamefont {Passerini}}, \bibinfo {author} {\bibfnamefont
  {Y.}~\bibnamefont {Fedoryshyn}},\ and\ \bibinfo {author} {\bibfnamefont
  {J.}~\bibnamefont {Leuthold}},\ }\bibfield  {title} {\bibinfo {title}
  {{Metamaterial graphene photodetector with bandwidth exceeding 500
  gigahertz}},\ }\href {https://doi.org/10.1126/science.adg8017} {\bibfield
  {journal} {\bibinfo  {journal} {Science}\ }\textbf {\bibinfo {volume}
  {380}},\ \bibinfo {pages} {1169} (\bibinfo {year} {2023})}\BibitemShut
  {NoStop}%
\bibitem [{\citenamefont {Wei}\ \emph {et~al.}(2021)\citenamefont {Wei},
  \citenamefont {Xu}, \citenamefont {Dong}, \citenamefont {Qiu},\ and\
  \citenamefont {Lee}}]{Wei2021}%
  \BibitemOpen
  \bibfield  {author} {\bibinfo {author} {\bibfnamefont {J.}~\bibnamefont
  {Wei}}, \bibinfo {author} {\bibfnamefont {C.}~\bibnamefont {Xu}}, \bibinfo
  {author} {\bibfnamefont {B.}~\bibnamefont {Dong}}, \bibinfo {author}
  {\bibfnamefont {C.~W.}\ \bibnamefont {Qiu}},\ and\ \bibinfo {author}
  {\bibfnamefont {C.}~\bibnamefont {Lee}},\ }\bibfield  {title} {\bibinfo
  {title} {{Mid-infrared semimetal polarization detectors with configurable
  polarity transition}},\ }\href {https://doi.org/10.1038/s41566-021-00819-6}
  {\bibfield  {journal} {\bibinfo  {journal} {Nature Photonics}\ }\textbf
  {\bibinfo {volume} {15}},\ \bibinfo {pages} {614} (\bibinfo {year}
  {2021})}\BibitemShut {NoStop}%
\bibitem [{\citenamefont {Otteneder}\ \emph {et~al.}(2020)\citenamefont
  {Otteneder}, \citenamefont {Hubmann}, \citenamefont {Lu}, \citenamefont
  {Kozlov}, \citenamefont {Golub}, \citenamefont {Watanabe}, \citenamefont
  {Taniguchi}, \citenamefont {Efetov},\ and\ \citenamefont
  {Ganichev}}]{Otteneder2020}%
  \BibitemOpen
  \bibfield  {author} {\bibinfo {author} {\bibfnamefont {M.}~\bibnamefont
  {Otteneder}}, \bibinfo {author} {\bibfnamefont {S.}~\bibnamefont {Hubmann}},
  \bibinfo {author} {\bibfnamefont {X.}~\bibnamefont {Lu}}, \bibinfo {author}
  {\bibfnamefont {D.~A.}\ \bibnamefont {Kozlov}}, \bibinfo {author}
  {\bibfnamefont {L.~E.}\ \bibnamefont {Golub}}, \bibinfo {author}
  {\bibfnamefont {K.}~\bibnamefont {Watanabe}}, \bibinfo {author}
  {\bibfnamefont {T.}~\bibnamefont {Taniguchi}}, \bibinfo {author}
  {\bibfnamefont {D.~K.}\ \bibnamefont {Efetov}},\ and\ \bibinfo {author}
  {\bibfnamefont {S.~D.}\ \bibnamefont {Ganichev}},\ }\bibfield  {title}
  {\bibinfo {title} {{Terahertz Photogalvanics in Twisted Bilayer Graphene
  Close to the Second Magic Angle}},\ }\href
  {https://doi.org/10.1021/acs.nanolett.0c02474} {\bibfield  {journal}
  {\bibinfo  {journal} {Nano Letters}\ }\textbf {\bibinfo {volume} {20}},\
  \bibinfo {pages} {7152} (\bibinfo {year} {2020})},\ \Eprint
  {https://arxiv.org/abs/2006.08324} {arXiv:2006.08324} \BibitemShut {NoStop}%
\bibitem [{\citenamefont {Karch}\ \emph {et~al.}(2010)\citenamefont {Karch},
  \citenamefont {Olbrich}, \citenamefont {Schmalzbauer}, \citenamefont {Zoth},
  \citenamefont {Brinsteiner}, \citenamefont {Fehrenbacher}, \citenamefont
  {Wurstbauer}, \citenamefont {Glazov}, \citenamefont {Tarasenko},
  \citenamefont {Ivchenko}, \citenamefont {Weiss}, \citenamefont {Eroms},
  \citenamefont {Yakimova}, \citenamefont {Lara-Avila}, \citenamefont
  {Kubatkin},\ and\ \citenamefont {Ganichev}}]{Karch2010}%
  \BibitemOpen
  \bibfield  {author} {\bibinfo {author} {\bibfnamefont {J.}~\bibnamefont
  {Karch}}, \bibinfo {author} {\bibfnamefont {P.}~\bibnamefont {Olbrich}},
  \bibinfo {author} {\bibfnamefont {M.}~\bibnamefont {Schmalzbauer}}, \bibinfo
  {author} {\bibfnamefont {C.}~\bibnamefont {Zoth}}, \bibinfo {author}
  {\bibfnamefont {C.}~\bibnamefont {Brinsteiner}}, \bibinfo {author}
  {\bibfnamefont {M.}~\bibnamefont {Fehrenbacher}}, \bibinfo {author}
  {\bibfnamefont {U.}~\bibnamefont {Wurstbauer}}, \bibinfo {author}
  {\bibfnamefont {M.~M.}\ \bibnamefont {Glazov}}, \bibinfo {author}
  {\bibfnamefont {S.~A.}\ \bibnamefont {Tarasenko}}, \bibinfo {author}
  {\bibfnamefont {E.~L.}\ \bibnamefont {Ivchenko}}, \bibinfo {author}
  {\bibfnamefont {D.}~\bibnamefont {Weiss}}, \bibinfo {author} {\bibfnamefont
  {J.}~\bibnamefont {Eroms}}, \bibinfo {author} {\bibfnamefont
  {R.}~\bibnamefont {Yakimova}}, \bibinfo {author} {\bibfnamefont
  {S.}~\bibnamefont {Lara-Avila}}, \bibinfo {author} {\bibfnamefont
  {S.}~\bibnamefont {Kubatkin}},\ and\ \bibinfo {author} {\bibfnamefont
  {S.~D.}\ \bibnamefont {Ganichev}},\ }\bibfield  {title} {\bibinfo {title}
  {{Dynamic Hall Effect Driven by Circularly Polarized Light in a Graphene
  Layer}},\ }\href {https://doi.org/10.1103/PhysRevLett.105.227402} {\bibfield
  {journal} {\bibinfo  {journal} {Physical Review Letters}\ }\textbf {\bibinfo
  {volume} {105}},\ \bibinfo {pages} {227402} (\bibinfo {year}
  {2010})}\BibitemShut {NoStop}%
\bibitem [{\citenamefont {Dyakonov}\ and\ \citenamefont
  {Shur}(1996)}]{Dyakonov1996}%
  \BibitemOpen
  \bibfield  {author} {\bibinfo {author} {\bibfnamefont {M.}~\bibnamefont
  {Dyakonov}}\ and\ \bibinfo {author} {\bibfnamefont {M.}~\bibnamefont
  {Shur}},\ }\bibfield  {title} {\bibinfo {title} {{Detection, mixing, and
  frequency multiplication of terahertz radiation by two-dimensional electronic
  fluid}},\ }\href {https://doi.org/10.1109/16.485650} {\bibfield  {journal}
  {\bibinfo  {journal} {IEEE Transactions on Electron Devices}\ }\textbf
  {\bibinfo {volume} {43}},\ \bibinfo {pages} {380} (\bibinfo {year}
  {1996})}\BibitemShut {NoStop}%
\bibitem [{\citenamefont {Ryzhii}\ and\ \citenamefont
  {Shur}(2006)}]{Ryzhii2006}%
  \BibitemOpen
  \bibfield  {author} {\bibinfo {author} {\bibfnamefont {V.}~\bibnamefont
  {Ryzhii}}\ and\ \bibinfo {author} {\bibfnamefont {M.~S.}\ \bibnamefont
  {Shur}},\ }\bibfield  {title} {\bibinfo {title} {{Resonant terahertz detector
  utilizing plasma oscillations in two-dimensional electron system with lateral
  Schottky junction}},\ }\href {https://doi.org/10.1143/JJAP.45.L1118}
  {\bibfield  {journal} {\bibinfo  {journal} {Japanese Journal of Applied
  Physics, Part 2: Letters}\ }\textbf {\bibinfo {volume} {45}},\ \bibinfo
  {pages} {8} (\bibinfo {year} {2006})}\BibitemShut {NoStop}%
\bibitem [{\citenamefont {Svintsov}(2018)}]{Svintsov2018b}%
  \BibitemOpen
  \bibfield  {author} {\bibinfo {author} {\bibfnamefont {D.}~\bibnamefont
  {Svintsov}},\ }\bibfield  {title} {\bibinfo {title} {{Exact Solution for
  Driven Oscillations in Plasmonic Field-Effect Transistors}},\ }\href
  {https://doi.org/10.1103/PhysRevApplied.10.024037} {\bibfield  {journal}
  {\bibinfo  {journal} {Physical Review Applied}\ }\textbf {\bibinfo {volume}
  {10}},\ \bibinfo {pages} {024037} (\bibinfo {year} {2018})}\BibitemShut
  {NoStop}%
\bibitem [{\citenamefont {Ryzhii}\ and\ \citenamefont
  {Khrenov}(1995)}]{Ryzhii1995}%
  \BibitemOpen
  \bibfield  {author} {\bibinfo {author} {\bibfnamefont {V.}~\bibnamefont
  {Ryzhii}}\ and\ \bibinfo {author} {\bibfnamefont {G.}~\bibnamefont
  {Khrenov}},\ }\bibfield  {title} {\bibinfo {title} {{High-Frequency Operation
  of Lateral Hot-Electron Transistors}},\ }\href
  {https://doi.org/10.1109/16.370021} {\bibfield  {journal} {\bibinfo
  {journal} {IEEE Transactions on Electron Devices}\ }\textbf {\bibinfo
  {volume} {42}},\ \bibinfo {pages} {166} (\bibinfo {year} {1995})}\BibitemShut
  {NoStop}%
\end{thebibliography}%

\end{document}